# Combined tests based on restricted mean time lost for competing risks data[§]

**Jingjing Lyu**[1], **Yawen Hou**[2*], **Zheng Chen**[1*]

[1] Department of Biostatistics, Southern Medical University, Guangzhou, China

[2] Department of Statistics, Jinan University, Guangzhou, China

*Corresponding author: Yawen Hou and Zheng Chen

Sep. 2020

**Abstract:** Competing risks data are common in medical studies, and the sub-distribution hazard (SDH) ratio is considered an appropriate measure. However, because the limitations of hazard itself are not easy to interpret clinically and because the SDH ratio is valid only under the proportional SDH assumption, this article introduced an alternative index under competing risks, named restricted mean time lost (RMTL). Several test procedures were also constructed based on RMTL. First, we introduced the definition and estimation of RMTL based on Aalen-Johansen cumulative incidence functions. Then, we considered several combined tests based on the SDH and the RMTL difference (RMTLd). The statistical properties of the methods are evaluated using simulations and are applied to two examples. The type I errors of combined tests are close to the nominal level. All combined tests show acceptable power in all situations. In conclusion, RMTL can meaningfully summarize treatment effects for clinical decision making, and three combined tests have robust power under various conditions, which can be considered for statistical inference in real data analysis.

---

# 1 Introduction

During follow-up in a clinical study, patients may experience failure from multiple causes. The primarily observed or researched endpoint is recorded as the event of interest, while other endpoints that may preclude the occurrence of an event of interest are recorded as competing events [1, 2]. For example, in a study of H7N9 avian influenza virus (H7N9), patients may experience two main types of failure: death or cure [3]. If a researcher's endpoint of interest is cure, then death is considered a competing event. When competing risks exist, the risk of an event is overestimated by analysis methods based on a single endpoint, which may cause large errors and lead to incorrect conclusions [4,5]. One quantity of interest for competing risks data is the cumulative incidence function (CIF) [6,7], and a common method used to compare efficacy between groups is the Gray test (Gray) [8], for which the corresponding index is the sub-distribution hazard (SDH) ratio. However, the SDH ratio (SHR) has the following limitations [9-12]: i) the SHR has optimal power only under the proportional SDH assumption; ii) the SDH function is derived as $\lambda(t) = \lim_{\Delta t \to 0} \frac{P(t \leq T < t + \Delta t, \text{event } j \mid T > t \cup (T < t \cap \text{not event } j))}{\Delta t}$, which is based on the conditional probability. An SHR may complicate the interpretation of survival outcomes and may not meaningfully summarize treatment effects for clinical decision-making.

One solution is to set a restricted time point $\tau$ ($\tau > 0$) and obtain the restricted mean survival time (RMST) [13-15], which is the mean survival time during the time period 0 to $\tau$. However, the existing RMST methods are primarily based on single-event survival analysis. Calkins et al. [16] and Wang et al. [17] referred to the concept of RMST under competing risks. Calkins et al. [16] focused on the RMST spent in a state free of all events and obtained the corresponding descriptive statistics, while Wang et al. [17] built a regression model based on pseudovalues of RMST under competing risks. Andersen [18] referred to the expected value of patients' life years within the definition of the number of life years lost under competing risks setting (NLYL) and proposed a regression model based on the NLYL pseudovalues observation.

Zhao et al. [12] introduced NLYL by using real data and renamed NLYL to restricted mean time lost (RMTL), which corresponds to the area under CIF curves. Lyu et al. [19] developed two hypothesis tests based on RMTL difference (RMTLd). However, the two tests maintain stable type I error rates and high powers in limited situations.

The SDH-based test, Gray, could maintain a stable type I error rate but needs to satisfy the proportional SDH assumption when using. RMTL is a great alternative index of SDH, but the existing RMTL-based tests [19] are also limited. Thus, we consider an alternative type of the combined tests with a combination of SDH and RMTL, which are available to increase the robustness of either of the primary tests [20]. The structure of this article is as follows. In Section 2, we introduce the definition and estimation of the RMTL under competing risks and statistical inference methods. Simulation studies are reported in Section 3. Furthermore, we provide two examples for a detailed understanding in Section 4. We finish with a discussion of our research.

## 2 Methods

Assume a randomized study of two groups (one control group for $g=1$ and one treatment group for $g=2$) with $n$ patients. The time-to-event is denoted by $T$, which is assumed to be independent of the censoring time $C$. Let $\tau$ denotes the cut-off point, which is less than or equal to the maximum observed follow-up time. Here, we choose one of the most commonly used options, the minimum time of the last event of interest in two groups, as $\tau$ [21,22]. Let $j$ denotes the event type, and without loss of generality, we assume one event of interest ($j=1$) and one competing event ($j=2$). The subsurvival function is $S_j(t) = P(T > t, j)$, and the subdensity function is $f_j(t) = -dS_j(t)/dt$. The CIF under time point $t$ of event $j$ is denoted by $I_j(t) = P(T \leq t, j)$. $I_1(t)$ and $I_2(t)$ are the CIFs for the event of interest and the competing event, respectively. Based on the nonparametric maximum likelihood estimation of the CIF, the Aalen-Johansen estimate [23] of $I_j(t)$ is $\hat{I}_j(t) = \sum_{t_i \leq t}(d_{ij}/n_i)\hat{S}(t_{i-1})$, where $d_{ij}$ is the number of events of type $j$ that occur at time $t_i$, $n_i$ is the number at risk at $t_i$, and $\hat{S}(t)$ is the Kaplan-Meier estimate when all events (both $j=1$ and $j=2$) are considered.

## 2.1 Estimation of Restricted Mean Time Lost in Competing Risks

For simplicity, we only considered the event of interest ($j$=1); then, the RMTL is given by [16,18-19]

$$\text{RMTL} = \int_0^\tau I_1(t)dt, \qquad (1)$$

which means the lifetime that a patient delays to occur the event of interest. Based on the CIF estimation of the event of interest, $\hat{I}_1(t)$, the estimate of RMTL is given by

$$\widehat{\text{RMTL}} = \int_0^\tau \hat{I}_1(t)dt.$$

The estimated variance of $\widehat{\text{RMTL}}$ [19] is

$$\widehat{\text{var}}_{\text{RMTL}} = 2\tau \int_0^\tau \hat{I}_1(t)dt - 2\int_0^\tau t\hat{I}_1(t)dt - [\int_0^\tau \hat{I}_1(t)dt]^2.$$

## 2.2 Hypothesis Tests

### 2.2.1 Basic Difference Test

Lyu et al [19] constructed test statistics based on the RMTLd between two groups, $\Delta_1(=X_1-X_2)$, where $X_g$ is the RMTL estimate of the $g$th group. This method is called the basic difference (Diff) test. Assume that the sample sizes in the two groups are $n_1$ and $n_2$. Then, the difference estimate is $\hat{\Delta}_1 = \hat{X}_1 - \hat{X}_2$, and the corresponding variance of $\hat{\Delta}_1$,

$$\widehat{\text{var}}(\hat{\Delta}_1) = \widehat{\text{var}}(\hat{X}_1)/n_1 + \widehat{\text{var}}(\hat{X}_2)/n_2$$

is derived by the delta method.

Under the null hypothesis $H_0: \Delta_1 = 0$, the test statistic is

$$Z_1 = \frac{\hat{\Delta}_1}{\sqrt{\widehat{\text{var}(\hat{\Delta}_1)}}},$$

which against a normal reference distribution in large samples.

### *2.2.2 Combined Test*

Some researchers [24-26] noted that the combination of two (or more) test statistics can increase the power compared with the use of a single test. Because the Gray test may lose power when non-proportional SDHs are present, we combined the Gray test and Diff test to maintain power under various scenarios.

### 一 *Minimum P-value Test*

First, we consider the minimum $P$-value test, which is constructed by

$$P_{\min P} = \min(p_{Gray}, p_{RMTLD}),$$

where $p_{Gray}$ is the model-based $P$-value of the Gray test, and $p_{RMTLD}$ is the model-based $P$-value of the Diff test.

$P_{\min P}$ follows a $Beta(1,2)$ distribution when $p_{Gray}$ and $p_{RMTLD}$ are independent of each other [26]. However, when $p_{Gray}$ and $p_{RMTLD}$ are positively correlated, $P_{\min P}$ no longer follows the above distribution. We denote the minimum $P$-value combination of the Gray test and the Diff test as PComb.

### 一 *Fisher's Combination Test*

In addition, we consider the combining function of Fisher's combination test, the test statistics are built by

$$P_{FCT} = -2(\ln(p_{Gray}) + \ln(p_{RMTLD})).$$

$P_{FCT}$ follows a $\chi^2(4)$ distribution when $p_{Gray}$ and $p_{RMDL}$ are independent of each other [26]. However, $P_{FCT}$ no longer follows the above distribution because $p_{Gray}$ and $p_{RMTLD}$ are positively correlated. We denote the Fisher's combination of the Gray test and Diff test as FComb.

## 一 Two-stage Test

In addition to combining *P*-values, we consider the two-stage procedure, which was introduced by Qiu and Sheng [27] and can obtain higher power under various scenarios. As the term suggests, two-stage means that two different methods are selected for testing in two different stages. For example, we choose *Test*1 and *Test*2 for testing. We use *Test*1 in stage one. If the null hypothesis is rejected, then the results are output; otherwise, we perform *Test*2 in stage two and output the results of *Test*2. *Test*1 has some limitations but is often powerful in certain situations. Therefore, we choose the Gray test, which needs to satisfy the proportional SDH assumption, for stage one, and we select the Diff test for stage two.

To fix the overall significance level at $\alpha$, we adjusted the significance levels of the two stages, denoted by $\alpha_1$ and $\alpha_2$, respectively. Let

$$\alpha_1 + P(reject\ B\ |\ not\ reject\ A) \cdot (1-\alpha_1) = \alpha,$$

and $P(reject\ B\ |\ not\ reject\ A) = \alpha_2$. Without any prior information, we assumed that the results obtained in the two stages are independent. Then, we suppose $\alpha_1 = \alpha_2$ and write the following formula:

$$\alpha_1 + \alpha_1(1-\alpha_1) = \alpha.$$

That is,

$$\alpha_1 = 1 - \sqrt{1-\alpha}.$$

However, in practical applications, the results of stage one and stage two are strongly correlated, and $\alpha_2$ is difficult to solve. In addition, to avoid two stages that are too radical, we use a permutation test to resample the results and obtain the estimator of $P(reject\ B\ |\ not\ reject\ A)$. The final result is

$$P = \begin{cases} p_{Gray}, & p_1 \leq \alpha_1 \\ \alpha_1 + P(reject\ B\ |\ not\ reject\ A) \cdot (1-\alpha_1), & others \end{cases},$$

where $p_1$ is the computed P-value in stage one. We denote the combination of Gray test and Diff test as TComb.

## 3 Results

### 3.1 Simulation Design

To compare the performance of the above tests with that of the Gray test, we conducted Monte Carlo simulations for the following situations (Figure 1): (A) two groups with the same CIF; (B) two groups with proportional SDH; (C) a late difference in the CIFs; and (D) an early difference in the CIFs. Both groups were set to have the same censoring rates of approximately 0, 15%, 30%, 45% and 60%. We also considered equal group sizes ($n_1=n_2=50, 100, 150$) and unequal group sizes ($n_1=50$, $n_2=100$; $n_1=50$, $n_2=150$; $n_1=50$, $n_2=200$). All simulations were performed using 5000 iterations and 200 internal iterations in permutations. The nominal significance level of each method was fixed at the conventional level of 0.05.

To investigate type I error under situation A (Figure 1A), the failure times of the two groups were generated from the CIF $I_1(t) = p_1(1-e^{-t})$ for event 1 (the event of interest) and $I_2(t) = (1-p_1)(1-e^{-t})$ for event 2, where $p_1 = I_1(\infty) = 0.7$. Therefore, the maximum cumulative incidence of event 1 was set to 70%. The power was evaluated using three alternative scenarios (Figure 1B-1D). In situation B (Figure 1B), two groups of failure times $T$ were generated from the CIFs $I_1(t) = 1-[1-p_1(1-e^{-t})]^{\exp\{\beta Z\}}$ and $I_2(t) = (1-p_1)^{\exp\{\beta Z\}}(1-e^{-t\exp\{\beta Z\}})$, where $Z=0$ and $Z=1$ in group 1 and group 2, respectively, and $p_1 = I_1(\infty) = 0.7$. Situation C (Figure 1C) and situation D (Figure 1D) were generated from CIFs with piecewise Weibull distributions of the form $I_1(t) = p_1\{1-\exp(-(t/2)^A)\}$ and $I_2(t) = (1-p_1)\{1-\exp(-(t/2)^A)\}$ for both groups, where the pieces were set at two years ($t=2$) to ensure a sufficient number of events before and after $t$. In situation C, the CIFs were generated with $A=2$ for both groups in $t \leq 2$. For $t > 2$, we set $A = 0.1$ and $A = 4$ for the two groups. In situation D, the CIFs were generated with $A = 0.1$ and $A = 4$ for the two groups in $t \leq 2$. For $t > 2$, we set $A = 2$ for both groups. The distribution of event 1 was based on the binomial distribution $B(N, p_1)$, where $N$ represents the sample size in each group. The censored times $C$ in the two groups were generated from uniform distributions $U(0, a)$ and $U(0, b)$. Varying the values of $a$ and $b$ results in different censoring. Then, each individual was assigned an observed time $t=\min(T, C)$, and the event indicator $\delta=IF[T \leq C]$, where $IF(\cdot)$ denotes the indicator function.

To summarize the simulation results, we applied analysis of variance (ANOVA) technique [28] to evaluate the type I error and power (see details in Appendix 1).

### 3.2 Simulation Results

*Situation A.* The reasonable range of the type I error rates (0.0440, 0.0560) is based on the formula $\alpha \pm 1.96 \times \sqrt{0.05(1-0.05)/5000}$ when the true error rate is at the nominal level (0.05) and 5000 replicated samples are considered [11]. As shown in Table 1, Diff are inflated and above normal in most situations. The three combined tests are the most stable among all methods. For all combinations of sample sizes and censoring rates, the average deviation from the nominal 5 percent level (Table A1 in Appendix 1) indicates that all the methods, except Diff, are within a reasonable range.

*Situation B.* As shown in Table 2, when two CIF curves have proportional SDH (Figure 1B), the powers of all the tests increase with increasing sample size but decline with increasing censoring rate: when the censoring increases to 60%, the power of all tests sharply decreases. As shown in Table 2 and Table A2 (Appendix 1), the Gray test has the highest power, and all the other tests also have acceptable power (approximately over 80%).

*Situation C.* When considering the late difference in the CIF curves (Figure 1C), Table 3 demonstrates that the powers of all the tests decline with increasing censoring rate. As shown in Table 3 and Table A2 (Appendix 1), Gray, TComb and PComb demonstrate optimal power in this situation, whereas the other tests maintain relatively low power.

*Situation D.* For the early difference in the CIF curves (Figure 1D), Table 4 summarizes the results of the power estimates. All tests exhibit gradually increasing power with increasing censoring. The two CIF curves are not convergent in the later period but are divergent with the increased censoring, which makes the increased difference between the two CIF curves proportional. According to Table 4 and Table A2 (Appendix 1), all tests, except the Gray test, maintain a certain power.

In general, according to the simulation results shown in Tables 1-4 and Tables A1-A2 (Appendix 1), the Gray test maintains relatively high power in the case of proportional SDH (situation B), while three combined tests perform better in the early difference case (situation D). Considering all situations, combinations of sample sizes, and censoring rates, TComb has the highest power, followed by PComb.

## 4 Applications

In this section, we apply the above tests to two data examples. Furthermore, similar to Calkins et al. [16], we propose three test procedures on RMST under competing risks (RC), which focused more on composite events in competing risks settings. We denote the corresponding test as Diff$^*$ (see details and the simulation results in Appendix 2). In addition, we consider the RMST, which is based on a single-event survival analysis with the event of interest (RMSTi) and the composite event (RMSTc), respectively.

### 4.1 Example 1: The Proportional Subdistribution Hazards Assumption is Satisfied

A study of H7N9 avian influenza virus (H7N9) included 103 patients [3] divided into a young and middle-aged group (younger than 60 years of age) of 46 cases and an old group (60 years old and above) of 57 cases. The censoring rates in the two groups were approximately 50% and 35%, respectively. The study included two types of events, of which death due to the event of interest and cure was set as competing events. At the end of follow-up, 23 patients (7 from the young and middle-aged group and 16 from the old group) experienced an event of interest, and 30 patients experienced competing events (16 from the young and middle-aged group and 14 from the old group). Meanwhile, a test of the proportional SDH assumption [29] yielded a result of *P*=0.681.

Figure 2A shows the cumulative incidence curves for death. According to Table 5, the estimated SHR is 1.98 (95% CI: 0.83, 4.77) in favor of the young and middle-aged group, and the Gray test yielded a result of *P*=0.121. Figure 2B and Figure 2C show the CIF for death and the complement of the CIF for cure at the restricted time point of $\tau = 34$ days among different groups. Some descriptive results are shown in Table 5. The RMTL for the young and middle-aged group (Figure 2B, area 'S1') is 2.46 days (95% CI: 0.42, 4.50) and that of the old group (Figure 2C, area 'S1') is 5.02 days (95% CI: 2.67, 7.37), which means that the young and middle-aged group experienced a 2.56-day-delay (95% CI: -0.55, 5.68) in time to death compared with the old group. Additionally, RMSTi provides an estimation of -2.58 days (95% CI: -5.85, 0.69). As shown in Table 6, the *P*-value of TComb is 0.067, which is the minimum value among all test procedures.

In addition, we consider the composite endpoint of death and cure, and the results are shown in Tables 5 and 6. The RC for the young and middle-aged group (Figure 2B, area 'S2') is 25.38 days (95% CI: 22.41, 28.35) and that of the old group (Figure 2C, area 'S2') is 26.23 days (95% CI: 23.82, 28.64). Therefore, the RC for the old group experiences an advance of 0.85 days (95% CI: -2.98, 4.67) compared with the young and middle-aged group. Additionally, RMSTc provides an estimation of 0.85 days (*P*=0.663).

**4.2 Example 2: The Proportional Subdistribution Hazards Assumption is Violated**

A previous study compared the effects of radiotherapy in the treatment of patients with lymphocytic leukemia (LL). We randomly extracted 1400 patients from the Surveillance, Epidemiology, End Results (SEER) Program. Among these patients, 82 patients who were treated with radiotherapy were included in the "radiotherapy" group, and 1318 patients who were not were included in the "no radiotherapy" group. The censoring rates in the two groups were approximately 3% and 44%, respectively. We defined death from LL as the event of interest, and death from other causes was recorded as a competing event. At the end of follow-up, 364 patients (31 were treated with radiotherapy and 333 were not) experienced an event of interest, and 467 patients (15 were treated with radiotherapy and 452 were not) experienced competing events. A test of the proportional SDH assumption indicates that the assumption is indeed violated ($P$=0.006).

Figure 3A shows the cumulative incidence curves of death from LL. The estimated SHR of radiotherapy vs. no radiotherapy is 1.45 (95% CI: 0.98, 2.14) (Table 5), and the Gray test yields a result of $P$=0.072 (Table 6). Figures 3B and 3C show the CIF for death from LL and the complement of the CIF for death from other causes at the restricted time point $\tau=15.25$ years among different groups. According to the standard estimation, the RMTL for the no radiotherapy group (Figure 3B, area 'S1') is 2.96 years (95% CI: 2.69, 3.24) and that of the radiotherapy group (Figure 3C, area 'S1') is 4.68 years (95% CI: 3.34, 6.03), which means that patients with no radiotherapy experienced a 1.72-year-delay (95% CI: 0.35, 3.09) free from death to lymphocytic leukemia compared with radiotherapy patients. The RMSTi difference between groups is of -0.57 years (95% CI: -2.02, 0.87). As shown in Table 6, the $P$-values from all event-of-interest-related tests, except for Gray ($P$=0.072) and RMSTi ($P$=0.437), are less than 0.05.

In the analysis of the composite endpoint, the RC for patients with no radiotherapy (Figure 3B, area 'S2') is 8.52 years (95% CI: 8.21, 8.82) (Table 5) and that of the radiotherapy (Figure 3C, area 'S2') group is 9.09 years (95% CI: 7.70, 10.48), corresponding to a 0.57 year difference (95% CI: -0.85, 1.99) between the two groups. In addition, RMSTc shows a 0.57 year advance in the radiotherapy group, again with standard estimation of RC.

## 5 Discussion

In this paper, we introduced the index of RMTL in the competing risks setting. As the combined tests can increase the power compared with the use of a single test, we developed three statistical inference methods based on a combination of Gray and Diff tests. From our simulation results, all combined tests show an improvement in power with a stable type I error rate. Under non-proportional SDH, TComb and PComb achieved optimal power in the early difference case (situation D). Moreover, the Gray test maintained relatively high power when the two CIF curves had proportional SDH (situation B). In addition to RMTL, we also considered a composite event in competing risks setting, which is denoted as RC. The detailed procedures and simulations are shown in Appendix 2.

The quantification of treatment effects is complex in the presence of competing risks [12], and the common RMST methods, which are based on single-event survival analysis, cannot provide a precise solution. As shown in Table 5, RMSTi may not be applicable when competing risks appear because it may underestimate the RMST. Calkins et al. [16] described the RMST under competing risks settings based on the inverse-probability-weighted CIF and performed a descriptive analysis of real clinical data for composite events. However, when the defined events are opposites, such as death and cure in example 1 (H7N9), they cannot be simply combined into a composite endpoint [30]. Therefore, RC and RMSTc are not applicable in this situation. In example 2, both RMTL and RC are applicable. However, in example 2, RC, which is not significant at the nominal 0.05 level, may cover the effect of the event of interest, in contrast to RMTL. In addition, the point estimates of RC are equal to the RMSTc (Appendix 3), but the confidence interval of RC is somewhat narrower than that of RMSTc (Table 5). The most commonly used SDH-based method is not valid under non-proportional SDH and lacks of description in each group with only a summary measure of SHR. Without baseline information, SHR is not meaningful for clinical interpretation. Compared to that, RMTL is not based on any assumption and can directly reflect the survival of patients. Some problems remain to be solved. First, the variance of the Diff test is derived by the delta method [19]. However, from the simulations, we found the delta method is not the most appropriate method for estimating variance, which may cause deviation in the statistical distribution. In addition, many methods can be used to select the restricted time point $\tau$ under right-censored data conditions without competing risks [22], and different $\tau$ values may have various effects on the results [21,31]. For a more objective comparison of several methods, we chose the minimum of the last event of interest time as $\tau$, which is the most commonly used method. We will consider these issues in a future study.

The choice of primary summary measure for combined tests remains to be solved. When there is proportional SDH, we still recommend SHR to be reported as the primary measure; when the proportional SDH assumption is not satisfied, RMTLd could be a great alternative to SHR. However, when describing the treatment effect for each group, RMTL can always be a meaningful summarize for clinical interpretation. The proposed TComb test is robust and have wide applicability, which can be considered for statistical inference in real data analysis.

**Supporting information**

Additional supporting information (Appendix 1-3) may be found online in the Supporting Information section at the end of the article.

**Appendix 1:** Contains the theory of variance techniques (ANOVA) to evaluate the type I error and power and the results of RMTL by using ANOVA.

**Appendix 2:** Contains the method of restricted mean survival time under competing risks (RC) and the simulation results of RC.

**Appendix 3:** Contains the proof that RC is consistent with RMST based on composite events (RMSTc).

**FUNDING**

This work is supported by the National Natural Science Foundation of China (82173622, 81673268, 81903411), the Guangdong Basic and Applied Basic Research Foundation (2019A1515011506) and Natural Science Foundation of Guangdong Province (2018A030313849).

**Table 1** Type I error of the test procedures for situation A: two groups with the same CIF

| $n_1, n_2$ | Censoring rate | Gray | Diff | Combined tests | | |
|---|---|---|---|---|---|---|
| | | | | PComb | FComb | TComb |
| 50,50 | 0% | 0.0464 | 0.0534 | 0.0498 | 0.0474 | 0.0506 |
| | 15% | 0.0494 | 0.0634 | 0.0514 | 0.0496 | 0.0512 |
| | 30% | 0.0492 | 0.0744 | 0.0498 | 0.0512 | 0.0496 |
| | 45% | 0.0510 | 0.0782 | 0.0502 | 0.0536 | 0.0506 |
| | 60% | 0.0480 | 0.0788 | 0.0502 | 0.0494 | 0.0494 |
| 100,100 | 0% | 0.0556 | 0.0558 | 0.0536 | 0.0548 | 0.0528 |
| | 15% | 0.0516 | 0.0656 | 0.0524 | 0.0530 | 0.0534 |
| | 30% | 0.0518 | 0.0776 | 0.0528 | 0.0532 | 0.0534 |
| | 45% | 0.0502 | 0.0824 | 0.0522 | 0.0522 | 0.0526 |
| | 60% | 0.0534 | 0.0900 | 0.0518 | 0.0510 | 0.0518 |
| 150,150 | 0% | 0.0520 | 0.0546 | 0.0532 | 0.0550 | 0.0472 |
| | 15% | 0.0536 | 0.0646 | 0.0514 | 0.0536 | 0.0512 |
| | 30% | 0.0516 | 0.0722 | 0.0470 | 0.0484 | 0.0492 |
| | 45% | 0.0502 | 0.0772 | 0.0448 | 0.0472 | 0.0462 |
| | 60% | 0.0432 | 0.0774 | 0.0470 | 0.0442 | 0.0472 |
| 50,100 | 0% | 0.0554 | 0.0604 | 0.0528 | 0.0530 | 0.0528 |
| | 15% | 0.0578 | 0.0710 | 0.0566 | 0.0560 | 0.0580 |
| | 30% | 0.0536 | 0.0780 | 0.0498 | 0.0514 | 0.0522 |
| | 45% | 0.0498 | 0.0788 | 0.0500 | 0.0490 | 0.0514 |
| | 60% | 0.0468 | 0.0814 | 0.0482 | 0.0492 | 0.0480 |
| 50,150 | 0% | 0.0496 | 0.0560 | 0.0508 | 0.0506 | 0.0528 |
| | 15% | 0.0538 | 0.0686 | 0.0478 | 0.0484 | 0.0494 |
| | 30% | 0.0550 | 0.0782 | 0.0468 | 0.0482 | 0.0494 |
| | 45% | 0.0486 | 0.0776 | 0.0450 | 0.0458 | 0.0484 |
| | 60% | 0.0450 | 0.0822 | 0.0444 | 0.0422 | 0.0440 |
| 50,200 | 0% | 0.0526 | 0.0592 | 0.0520 | 0.0502 | 0.0544 |
| | 15% | 0.0562 | 0.0708 | 0.0526 | 0.0530 | 0.0552 |
| | 30% | 0.0544 | 0.0800 | 0.0512 | 0.0526 | 0.0540 |
| | 45% | 0.0512 | 0.0818 | 0.0510 | 0.0504 | 0.0494 |
| | 60% | 0.0436 | 0.0866 | 0.0472 | 0.0450 | 0.0476 |

**Table 2** Power of the test procedures for situation B: two groups with proportional subdistribution hazards

| $n_1, n_2$ | Censoring rate | Gray | Diff | Combined tests | | |
|---|---|---|---|---|---|---|
| | | | | PComb | FComb | TComb |
| 50,50 | 0% | 0.7950 | 0.8132 | 0.8002 | 0.7996 | 0.8082 |
| | 15% | 0.7836 | 0.7988 | 0.7676 | 0.7672 | 0.7732 |
| | 30% | 0.7072 | 0.7340 | 0.6718 | 0.6838 | 0.6748 |
| | 45% | 0.6350 | 0.6664 | 0.5878 | 0.6106 | 0.5974 |
| | 60% | 0.4686 | 0.4802 | 0.3886 | 0.4218 | 0.4062 |
| 100,100 | 0% | 0.9756 | 0.9818 | 0.9782 | 0.9778 | 0.9792 |
| | 15% | 0.9740 | 0.9796 | 0.9696 | 0.9710 | 0.9728 |
| | 30% | 0.9474 | 0.9598 | 0.9362 | 0.9404 | 0.9366 |
| | 45% | 0.9072 | 0.9260 | 0.8848 | 0.8960 | 0.8928 |
| | 60% | 0.7690 | 0.7806 | 0.6938 | 0.7188 | 0.7058 |
| 150,150 | 0% | 0.9974 | 0.9986 | 0.9986 | 0.9984 | 0.9984 |
| | 15% | 0.9976 | 0.9980 | 0.9972 | 0.9974 | 0.9974 |
| | 30% | 0.9944 | 0.9952 | 0.9914 | 0.9924 | 0.9918 |
| | 45% | 0.9842 | 0.9858 | 0.9788 | 0.9812 | 0.9788 |
| | 60% | 0.9090 | 0.9220 | 0.8630 | 0.8840 | 0.8792 |
| 50,100 | 0% | 0.9062 | 0.9162 | 0.9126 | 0.9108 | 0.9146 |
| | 15% | 0.9036 | 0.8958 | 0.8826 | 0.8860 | 0.8856 |
| | 30% | 0.8434 | 0.8474 | 0.7908 | 0.8094 | 0.8044 |
| | 45% | 0.7724 | 0.7718 | 0.7088 | 0.7294 | 0.7210 |
| | 60% | 0.5934 | 0.5610 | 0.4866 | 0.5202 | 0.5118 |
| 50,150 | 0% | 0.9276 | 0.9432 | 0.9392 | 0.9376 | 0.9414 |
| | 15% | 0.9292 | 0.9288 | 0.9214 | 0.9234 | 0.9220 |
| | 30% | 0.8788 | 0.8766 | 0.8388 | 0.8516 | 0.8490 |
| | 45% | 0.8172 | 0.8116 | 0.7508 | 0.7764 | 0.7632 |
| | 60% | 0.6500 | 0.5958 | 0.5292 | 0.5654 | 0.5634 |
| 50,200 | 0% | 0.9416 | 0.9564 | 0.9226 | 0.9212 | 0.9404 |
| | 15% | 0.9460 | 0.9458 | 0.9208 | 0.9220 | 0.9294 |
| | 30% | 0.9012 | 0.8986 | 0.8758 | 0.8828 | 0.8832 |
| | 45% | 0.8464 | 0.8346 | 0.8206 | 0.8308 | 0.8286 |
| | 60% | 0.6840 | 0.6278 | 0.6388 | 0.6636 | 0.6568 |

**Table 3** Power of the test procedures for situation C: two groups with a late difference

| $n_1, n_2$ | Censoring rate | Gray | Diff | Combined tests | | |
|---|---|---|---|---|---|---|
| | | | | PComb | FComb | TComb |
| 50,50 | 0% | 0.4614 | 0.1146 | 0.3984 | 0.2616 | 0.3546 |
| | 15% | 0.3946 | 0.1118 | 0.3168 | 0.2084 | 0.2904 |
| | 30% | 0.2910 | 0.1020 | 0.2126 | 0.1510 | 0.2068 |
| | 45% | 0.1558 | 0.0896 | 0.1072 | 0.0822 | 0.1144 |
| | 60% | 0.0618 | 0.0960 | 0.0506 | 0.0476 | 0.0550 |
| 100,100 | 0% | 0.7512 | 0.2380 | 0.7024 | 0.5246 | 0.6568 |
| | 15% | 0.6650 | 0.2228 | 0.6002 | 0.4468 | 0.5634 |
| | 30% | 0.5296 | 0.2004 | 0.4504 | 0.3390 | 0.4176 |
| | 45% | 0.3042 | 0.1418 | 0.2236 | 0.1696 | 0.2220 |
| | 60% | 0.0922 | 0.0974 | 0.0634 | 0.0588 | 0.0712 |
| 150,150 | 0% | 0.8952 | 0.3646 | 0.8670 | 0.7240 | 0.8382 |
| | 15% | 0.8364 | 0.3390 | 0.7846 | 0.6300 | 0.7638 |
| | 30% | 0.7050 | 0.2954 | 0.6236 | 0.4768 | 0.6070 |
| | 45% | 0.4178 | 0.2034 | 0.3064 | 0.2418 | 0.3230 |
| | 60% | 0.1204 | 0.1090 | 0.0786 | 0.0666 | 0.0952 |
| 50,100 | 0% | 0.5736 | 0.1406 | 0.5076 | 0.3438 | 0.4648 |
| | 15% | 0.4734 | 0.1352 | 0.4102 | 0.2764 | 0.3686 |
| | 30% | 0.3370 | 0.1214 | 0.2678 | 0.1908 | 0.2456 |
| | 45% | 0.1772 | 0.1002 | 0.1288 | 0.1012 | 0.1350 |
| | 60% | 0.0658 | 0.1026 | 0.0570 | 0.0474 | 0.0624 |
| 50,150 | 0% | 0.6294 | 0.2076 | 0.5704 | 0.4216 | 0.5220 |
| | 15% | 0.5902 | 0.1932 | 0.4946 | 0.3676 | 0.4834 |
| | 30% | 0.5010 | 0.1674 | 0.3816 | 0.2900 | 0.3936 |
| | 45% | 0.3030 | 0.1136 | 0.1816 | 0.1292 | 0.2154 |
| | 60% | 0.0884 | 0.0938 | 0.0466 | 0.0464 | 0.0616 |
| 50,200 | 0% | 0.6654 | 0.1514 | 0.5964 | 0.4564 | 0.5538 |
| | 15% | 0.5278 | 0.1452 | 0.5336 | 0.4106 | 0.5208 |
| | 30% | 0.3688 | 0.1288 | 0.4218 | 0.3200 | 0.4396 |
| | 45% | 0.1846 | 0.1092 | 0.1978 | 0.1468 | 0.2438 |
| | 60% | 0.0712 | 0.1098 | 0.0490 | 0.0524 | 0.0636 |

**Table 4** Power of the test procedures for situation D: two groups with an early difference

| $n_1, n_2$ | Censoring rate | Gray | Diff | Combined tests | | |
|---|---|---|---|---|---|---|
| | | | | PComb | FComb | TComb |
| 50,50 | 0% | 0.0760 | 0.2576 | 0.2068 | 0.1438 | 0.2268 |
| | 15% | 0.1010 | 0.3074 | 0.2272 | 0.1678 | 0.2398 |
| | 30% | 0.1418 | 0.3810 | 0.2806 | 0.2220 | 0.2850 |
| | 45% | 0.2140 | 0.5492 | 0.4136 | 0.3382 | 0.4154 |
| | 60% | 0.4038 | 0.8810 | 0.7740 | 0.6718 | 0.7692 |
| 100,100 | 0% | 0.1124 | 0.3698 | 0.3246 | 0.2198 | 0.3512 |
| | 15% | 0.1558 | 0.4230 | 0.3564 | 0.2652 | 0.3712 |
| | 30% | 0.2372 | 0.5132 | 0.4206 | 0.3426 | 0.4314 |
| | 45% | 0.3752 | 0.7110 | 0.5918 | 0.5156 | 0.5986 |
| | 60% | 0.6596 | 0.9710 | 0.9278 | 0.8840 | 0.9288 |
| 150,150 | 0% | 0.1360 | 0.4628 | 0.4190 | 0.2974 | 0.4436 |
| | 15% | 0.2096 | 0.5244 | 0.4512 | 0.3462 | 0.4682 |
| | 30% | 0.3298 | 0.6104 | 0.5130 | 0.4352 | 0.5276 |
| | 45% | 0.5100 | 0.8072 | 0.7144 | 0.6432 | 0.7196 |
| | 60% | 0.8330 | 0.9938 | 0.9748 | 0.9554 | 0.9772 |
| 50,100 | 0% | 0.0676 | 0.3276 | 0.2688 | 0.1772 | 0.2928 |
| | 15% | 0.0942 | 0.3804 | 0.2972 | 0.2044 | 0.3152 |
| | 30% | 0.1346 | 0.4716 | 0.3686 | 0.2732 | 0.3784 |
| | 45% | 0.2168 | 0.6586 | 0.5384 | 0.4308 | 0.5384 |
| | 60% | 0.4296 | 0.9486 | 0.8946 | 0.8112 | 0.8860 |
| 50,150 | 0% | 0.1246 | 0.2924 | 0.2422 | 0.1920 | 0.2644 |
| | 15% | 0.1754 | 0.3320 | 0.2596 | 0.2264 | 0.2694 |
| | 30% | 0.2520 | 0.4154 | 0.2986 | 0.2854 | 0.3022 |
| | 45% | 0.3766 | 0.6052 | 0.4460 | 0.4244 | 0.4506 |
| | 60% | 0.6460 | 0.9114 | 0.7742 | 0.7372 | 0.7682 |
| 50,200 | 0% | 0.0596 | 0.3956 | 0.2440 | 0.2010 | 0.2680 |
| | 15% | 0.0780 | 0.4502 | 0.2584 | 0.2310 | 0.2750 |
| | 30% | 0.1306 | 0.5540 | 0.3066 | 0.2932 | 0.3136 |
| | 45% | 0.2164 | 0.7352 | 0.4444 | 0.4290 | 0.4552 |
| | 60% | 0.4390 | 0.9760 | 0.7560 | 0.7346 | 0.7604 |

**Table 5** Descriptive statistics results of two examples

| | Example 1 | | | Example 2 | | |
|---|---|---|---|---|---|---|
| | Young and middle-aged (95% CI) | Old (95% CI) | Ratio/Difference[†] (95% CI) | No radiotherapy (95% CI) | Radiotherapy (95% CI) | Ratio/Difference[†] (95% CI) |
| SDH[¶] | \ | \ | 1.98 (0.83,4.77) | \ | \ | 1.45 (0.98,2.14) |
| RMTL[¶] | 2.46 (0.42,4.50) | 5.02 (2.67,7.37) | 2.56 (-0.55,5.68) | 2.96 (2.69,3.24) | 4.68 (3.34,6.03) | 1.72 (0.35,3.09) |
| RMSTi[¶] | 30.96 (28.72,33.20) | 28.38 (26.00,30.76) | -2.58 (-5.85,0.69) | 9.09 (7.68,10.49) | 8.52 (8.18,8.85) | -0.57 (-2.02,0.87) |
| RC[§] | 25.38 (22.41,28.35) | 26.23 (23.82,28.64) | 0.85 (-2.98,4.67) | 8.52 (8.21,8.82) | 9.09 (7.70,10.48) | 0.57 (-0.85, 1.99) |
| RMSTc[§] | 25.38 (22.41,28.35) | 26.23 (23.82,28.64) | 0.85 (-2.98,4.68) | 8.52 (8.18,8.85) | 9.09 (7.68,10.49) | 0.57 (-0.87,2.02) |

[¶]: is based on estimates of event of interest.

[§]: is based on estimates of composite events.

[†]: SDH is related to the SDH ratio; RMTL, RMSTi, RC, and RMSTc are related to the corresponding differences.

RMSTi: RMST based on the Kaplan-Meier method for event of interest.

RMSTc: RMST based on the Kaplan-Meier method for composite events.

**Table 6** Statistical inference results of the above tests for the two examples

| Method | P-value for example 1 | P-value for example 2 |
|---|---|---|
| Gray[¶] | 0.121 | 0.072 |
| Diff[¶] | 0.107 | 0.014 |
| PComb[¶] | 0.144 | 0.050 |
| FComb[¶] | 0.114 | 0.052 |
| TComb[¶] | 0.067 | 0.030 |
| RMSTi[¶] | 0.122 | 0.437 |
| Diff[*§] | 0.663 | 0.430 |
| RMSTc[§] | 0.663 | 0.430 |

[¶] : is based on estimates of event of interest.

[§] : is based on estimates of composite events.

RMSTi: RMST based on the Kaplan-Meier method for event of interest.

RMSTc: RMST based on the Kaplan-Meier method for composite events.

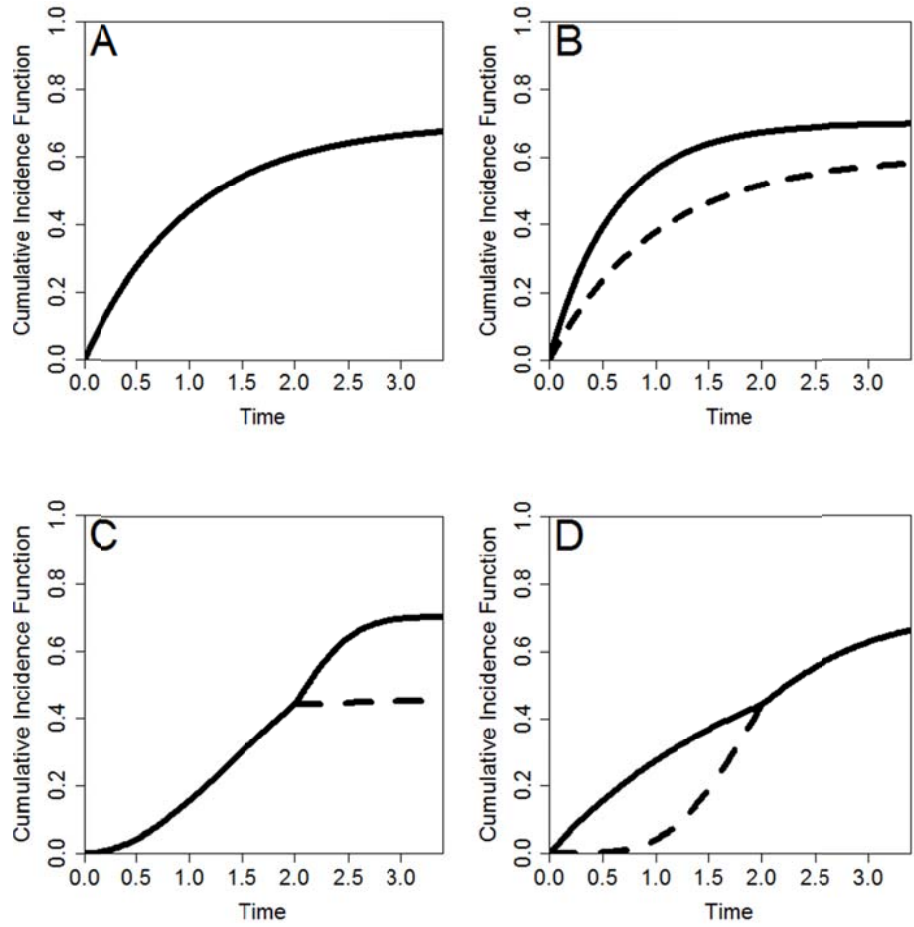

**Figure. 1** Four alternative scenarios in the simulation study—CIF curves of interest in four situations.

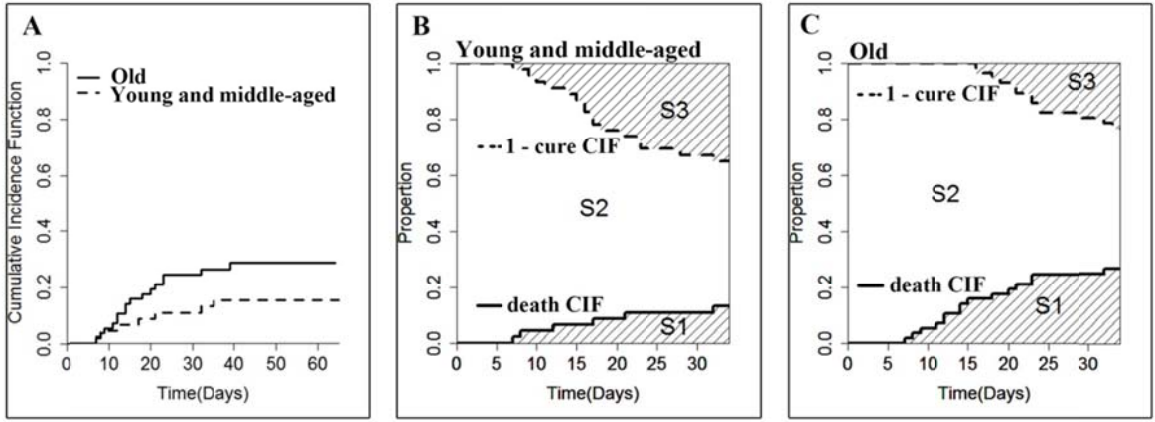

**Figure. 2** RMTL and RC estimates for Example 1 ($\tau$ =34 days). A) displays the cumulative incidence function (CIF) for death by age group. B) CIF for death (solid line) and 1-CIF for cure (dashed line) in the young and middle-aged group. C) CIF for death (solid line) and 1-CIF for cure (dashed line) in the old group.

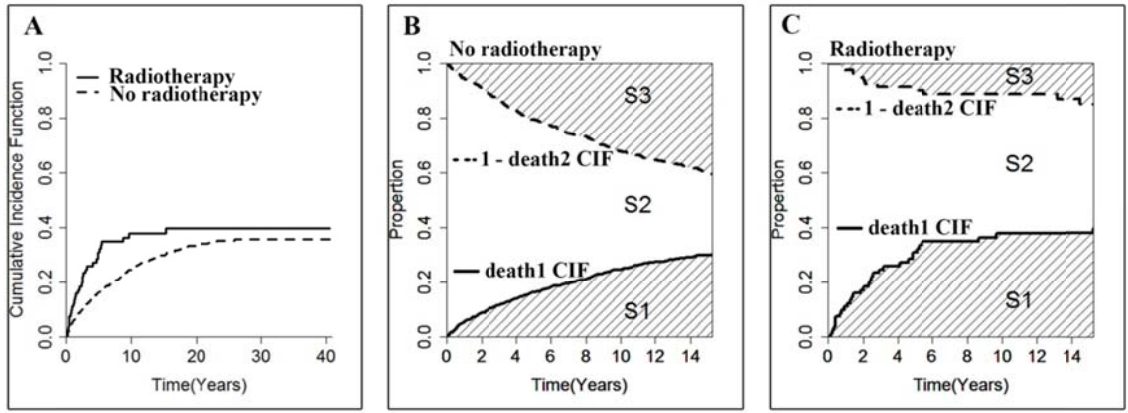

**Figure. 3** RMTL and RC estimates for Example 2 ($\tau$ =15.25 years). A) displays the cumulative incidence function (CIF) for death from lymphocytic leukemia (LL) by radiotherapy. B) CIF for death from LL (death1, solid line) and 1-CIF for death from other causes (1-death2, dashed line) in the no therapy group. C) CIF for death from LL (death1, solid line) and 1-CIF for death from other causes (1-death2, dashed line) in the therapy group.